\documentclass[reprint, amsmath, amssymb, superscriptaddress]{revtex4-2}
\usepackage{amsmath}
\usepackage{amsfonts}
\usepackage{graphicx}
\usepackage{hyperref}
\usepackage{booktabs}
\usepackage{float}
\usepackage{xcolor}
\usepackage{placeins}

\begin{document}
	
	\title{Towards Quantum Simulation of Rotating Nuclei using Quantum Variational Algorithms}
	\author{Dhritimalya Roy}
	\email{rdhritimalya@gmail.com}
	\affiliation{Department of Physics, Presidency University, Kolkata-700073, INDIA}
	\author{Somnath Nag}
	\email{somnathn.phy@itbhu.ac.in}
	\affiliation{Department of Physics, IIT-BHU, Varanasi, INDIA}
	
\begin{abstract}
	Quantum variational algorithms (QVAs) are increasingly potent tools for simulating quantum many-body systems on noisy intermediate-scale quantum (NISQ) devices. This work examines the application of the Variational Quantum Eigensolver (VQE) to four progressively complex models based on the cranked Nilsson-Strutinsky (CNS) framework. By incorporating single-particle spacings, pairing correlations, and rotational cranking terms, we evaluate VQE performance against exact diagonalization (ED) benchmarks. We provide a systematic benchmarking of VQE across a hierarchy of CNS-inspired Hamiltonians, explicitly identifying where hardware-efficient ansatz succeed and fail, and introducing quantum information diagnostics, the entanglement spectrum and Quantum Fisher Information, as novel probes of the pairing-rotation. Our results demonstrate that with a properly optimised multi-restart warm-starting strategy, VQE achieves near-machine-precision convergence ($|\Delta E| < 10^{-4}$) across the full cranking frequency range $\omega \in [0,1.2]$ and we confirm that the same strategy reproduces an established $^{6}$He shell-model pairing benchmark, demonstrating that the RealAmplitudes ansatz is expressively sufficient for this problem class.
The entanglement spectrum confirms the product-state character of the exact ground state throughout the pairing-rotation transition, while the Quantum Fisher Information identifies a finite-size precursor to the critical pair-breaking frequency. These results establish a systematic methodological baseline and provide a reproducible framework for the nuclear physics community.
\end{abstract}
	
	\maketitle
	
	\section{Introduction}
	In the age of noisy intermediate-scale quantum (NISQ) devices, quantum variational algorithms (QVAs) are becoming increasingly potent tools for simulating quantum systems.  These algorithms are especially well-suited for many-body problems where precise solutions are computationally unfeasible since they take advantage of the complementary advantages of quantum hardware and classical optimization.  In estimating ground-state energies, excited states, expectation values of observables, and other properties of complex quantum systems, the Variational Quantum Eigensolver (VQE), Variational Quantum Deflation (VQD), and ADAPT-VQE have demonstrated significant promise among these algorithms \cite{peru}-\cite{Grimsley:2018wnd}.
	
	Parallel to this, nuclear structure theory has long relied on the cranked Nilsson-Strutinsky (CNS) model \cite{ragnar,Juodagalvis:2005nj}.  It offers a useful framework for comprehending how fast rotating nuclei behave, allowing for in-depth understanding of shape transitions, band terminations, and collective phenomena at high angular momentum.  The CNS model, which has historically been handled through semi-classical and mean-field approximations, is still computationally demanding when it is expanded to incorporate pairing correlations, and angular momentum projections in full quantum many-body treatments.

In nuclear structural physics, the CNS framework itself has a long and illustrious history. The foundational description of nuclear rotation and deformation is laid out in the seminal works of Bohr and Mottelson \cite{BohrMottelson} and the comprehensive many-body treatment of Ring and Schuck \cite{RingSchuck}. The Nilsson model \cite{Nilsson:1955}, which underlies the single-particle level structure of the CNS approach, provides the microscopic basis for understanding deformed shell structure. The spontaneous breaking of rotational symmetry in rapidly rotating nuclei and its consequences for high-spin phenomenology are reviewed by Frauendorf \cite{Frauendorf:2001}. The competition between pairing correlations and rotational alignment which is the central physical theme of the present work has been studied extensively within mean-field frameworks \cite{Satula:1995, RingSchuck}, and the band termination sequences that result from the progressive alignment of valence nucleon angular momenta have been systematically documented and analysed \cite{Ragnarsson:1984, Afanasjev:1999}. These classical treatments, while powerful, are subject to well-known limitations: mean-field approximations break down near level crossings~\cite{Frauendorf:2001}, the quantum many-body sign problem afflicts Monte Carlo extensions~\cite{TroyerWiese2005}, and real-time rotational dynamics remain inaccessible to classical methods~\cite{Kashiwa:2018vxr,Gattringer:2016kco,Martinez:2016yna}. Quantum simulation offers a way  to address these limitations.
	
	With the advent of quantum computing technologies, various quantum many-body systems have served as benchmarks for testing the Variational Quantum Eigensolver (VQE), including models such as the Fermi-Hubbard, Ising, and Lipkin--Meshkov--Glick \cite{LMG1965}-\cite{Cervia:2020fkk}.  The potential of quantum simulation as an advancement in the field of nuclear physics has been studied by numerous researchers. The goal of these studies is to use quantum computing techniques to simulate intricate nuclear interactions, which are frequently unsolvable with traditional methods. Quantum simulators provide promising pathways to better understand the structure of atomic nuclei and the underlying dynamics of strong-force interactions by simulating nuclear systems at the quantum level, including few-body nuclei and nuclear matter. This expanding corpus of research highlights the applicability and viability of quantum simulation in resolving persistent issues in nuclear theory \cite{Dumitrescu:2018njn}-\cite{Stetcu:2021cbj}. Simulating CNS-like models using quantum algorithms opens up new possibilities to go beyond classical limitations, such as the sign problem and inability to incorporate real-time dynamics of the system \cite{Kashiwa:2018vxr}-\cite{Martinez:2016yna}. The ability to encode fermionic Hamiltonians and extract physically meaningful observables on quantum hardware offers a novel route to explore rotating nuclei with quantum-native methods.
	
	This work investigates the application of the VQE algorithm to a series of CNS-inspired Hamiltonians with increasing complexity. We begin with a minimal model suitable for current quantum hardware and progressively introduce more realistic physical ingredients, including pairing interactions, angular momentum operators, and particle number constraints. At each stage, quantum results are benchmarked against exact diagonalization (ED) to assess the fidelity of energy estimates, expectation values (such as $\langle J_x \rangle$), and entanglement entropy. This progressive refinement serves two purposes: (i) it improves the physical realism of the models, and (ii) it enables a systematic evaluation of how VQE performs under increasing representational demands. By exploring how well quantum variational simulations can replicate classical benchmarks across rotational frequencies and interaction strengths, we gain insight into both the strengths and limitations of current quantum strategies for nuclear structure modeling.

The primary novelty here is methodological, not phenomenological. The study is organized into the following major components: (i) systematic benchmarking of VQE across a hierarchy of CNS-inspired Hamiltonians of increasing complexity, explicitly mapping where hardware-efficient ansatz succeed and fail; (ii) the identification that optimizer strategy  rather than ansatz expressibility is the principal bottleneck for this problem class, demonstrated by achieving near-machine-precision convergence with a multi-restart warm-starting strategy; and (iii) the introduction of the entanglement spectrum and Quantum Fisher Information as quantum-information diagnostics of the pairing-rotation transition, providing physically interpretable quantities with no direct classical CNS analog. The Hamiltonians studied are intentionally schematic, following the long tradition of models such as the Lipkin--Meshkov--Glick\cite{LMG1965} and Richardson pairing models\cite{Richardson1963}, which use simplified parameters to capture universal physics. Accordingly, no quantitative correspondence with specific nuclear spectra is claimed; rather, the qualitative phenomenology that emerges is shown to be consistent with the universal physics of the CNS Hamiltonian class.
	
	Ultimately, this study demonstrates that quantum simulations, even on NISQ-era hardware or emulators, can capture non-trivial features of rotating nuclei. By starting from simple configurations and scaling up thoughtfully, our approach lays the groundwork for using quantum computing to address long-standing problems in the quantum many-body domain of nuclear physics, including those beyond the reach of classical methods.
	
	\section{Theoretical Models and Quantum Simulation Framework}
	To investigate the feasibility and accuracy of quantum variational algorithms in simulating nuclear rotational systems, we construct a sequence of schematic models inspired by the cranked Nilsson-Strutinsky (CNS) framework. Each model incorporates essential features of nuclear many-body physics, such as single-particle level structure, pairing correlations, and rotational cranking terms, while remaining tractable on near-term quantum simulators. The models are designed with increasing physical complexity, allowing for controlled benchmarking and stepwise validation of quantum algorithmic performance.
	
	In following  sections, we describe four progressively refined CNS-inspired Hamiltonians (Models I--IV). Each model is formulated in second quantization, mapped to qubit representations using the Jordan-Wigner transformation \cite{Kiss:2022kkz}-\cite{Stetcu:2021cbj},\cite{Seeley:2012zpt} and implemented within a VQE framework using suitable ansatz and optimizers. The choice of ansatz and level of circuit expressiveness are tailored to match the model's complexity, ensuring a balance between physical fidelity and hardware compatibility. The goal is to establish a scalable, modular simulation framework that can be incrementally improved toward realistic quantum simulations of nuclear structure.

	\section{Unified CNS-Inspired Hamiltonian and Qubit Mapping}
	\label{sec:ham}
	
	To establish a common framework for the four models studied in this work, we
	begin with a minimal fermionic Hamiltonian that captures the essential
	single-particle, pairing, and rotational features of the cranked
	Nilsson--Strutinsky (CNS) approach. We consider a small number of
	single-particle levels labelled by $i$, each with energy $\epsilon_i$, and
	associate with each level a pair of spin-orbitals $(i,\uparrow)$ and
	$(i,\downarrow)$. The fermionic creation and annihilation operators
	$a_{i\sigma}^\dagger$ and $a_{i\sigma}$ satisfy the standard anticommutation
	relations.
	
	\subsection{General Form of the Hamiltonian}
	The unified CNS-inspired Hamiltonian is written as
	\begin{equation}
		H(\omega)
		= \sum_{i,\sigma} \epsilon_i\, a_{i\sigma}^\dagger a_{i\sigma}
		- G \sum_{i} P_i^\dagger P_i
		- \omega \hat{J}_x,
		\label{eq:unified-H}
	\end{equation}
	where the terms represent:
	\begin{itemize}
		\item \emph{Single-particle structure}: 
		$n_{i\sigma} = a_{i\sigma}^\dagger a_{i\sigma}$ counts fermions in orbital
		$i$ with spin projection $\sigma$.
		\item \emph{Pairing interaction}: 
		$P_i^\dagger = a_{i\uparrow}^\dagger a_{i\downarrow}^\dagger$ creates a
		correlated pair in orbital $i$. The coupling strength $G>0$ promotes
		pair formation at low rotational frequencies.
		\item \emph{Cranking term}: 
		\(
		\hat{J}_x = \sum_{pq} \langle p | \hat{j}_x | q \rangle
		\, a_p^\dagger a_q
		\)
		models rotation about the intrinsic $x$ axis. The cranking frequency
		$\omega$ controls the degree of rotational alignment.
	\end{itemize}
	
	Although Eq.~(\ref{eq:unified-H}) is generic, it contains the essential
	ingredients of CNS phenomenology: single-particle shell structure, pair
	correlations, and the competition between pairing and rotational alignment as
	$\omega$ increases.
	
	\subsection{Choice of Numerical Parameters in the Models}
	The four models studied in this work instantiate Eq.~(\ref{eq:unified-H})
	using different numerical values for $\{\epsilon_i\}$, $G$, and the
	matrix elements of $\hat{J}_x$. These choices are intentionally schematic,
	allowing the models to illustrate CNS-like behaviour while remaining small
	enough for classical exact diagonalization.
	
	\begin{itemize}
		\item \textbf{Model I} employs a Pauli-word representation chosen to emulate
		a deformed mean field plus a simplified pairing structure. Here the values
		entering the coefficients of $ZIII$, $IZII$, and $IIZI$ mimic a split
		single-particle spectrum, while the $XXII$ and $YYII$ operators simulate
		an isovector pairing channel.
		\item \textbf{Model II} uses a four-level system with energies
		$\epsilon_i = \{-1.0, -0.5, 0.5, 1.0\}$, chosen symmetrically around zero.
		This reflects a typical situation in deformed oscillator potentials where
		orbitals occur in near-degenerate pairs. The pairing strength $G=0.5$ and
		a number-projection penalty ensure the two-particle sector is isolated.
		\item \textbf{Models III and IV} employ another commonly used schematic
		spectrum, $\epsilon_i = \{0.0, 0.2, 0.5, 0.8\}$, representing two spatial
		orbitals each split by spin. This choice generates a controlled, monotonic
		level spacing that is convenient for studying rotational alignment and
		the suppression of pairing with $\omega$. The pairing strength $G=0.6$ is
		selected to give a well-defined competition between pair condensation and
		Coriolis-induced pair breaking.
	\end{itemize}
	These parameter choices are not intended to reproduce detailed Nilsson spectroscopy. Instead, they are chosen so that the effects of pairing and rotation can be studied in a simple and stable setting, without additional complications from level crossings or strongly nucleus-specific shell structure. This approach follows the well-established tradition of schematic nuclear models, such as the Lipkin--Meshkov--Glick model\cite{LMG1965} and the Richardson pairing model\cite{Richardson1963}, which employ simplified parameters to illuminate universal physics rather than to reproduce the spectroscopy of specific nuclei. In this way, the models remain faithful to the CNS philosophy while also providing a useful testing ground for variational quantum algorithms applied to small interacting fermionic systems.
	
	\subsection{Mapping to Qubit Space}
	To apply quantum algorithms, the fermionic operators in
	Eq.~(\ref{eq:unified-H}) are mapped to qubit operators using the
	Jordan--Wigner (JW) transformation:
	\begin{align}
		a_p^\dagger &= \frac{1}{2}
		\left( X_p - i Y_p \right)
		\bigotimes_{q<p} Z_q,
		\\
		a_p &= \frac{1}{2}
		\left( X_p + i Y_p \right)
		\bigotimes_{q<p} Z_q,
	\end{align}
	where $p$ indexes the spin-orbitals $(i,\sigma)$.
	Under this mapping, number operators become $Z$ strings, pairing terms become
	products of $X$ and $Y$ operators, and $\hat{J}_x$ becomes a sparse sum of
	one-body qubit operators. The resulting qubit Hamiltonian takes the form
	\begin{equation}
		H_{\mathrm{JW}}(\omega)
		= \sum_{\alpha} c_\alpha\, P_\alpha,
	\end{equation}
	where $P_\alpha$ are Pauli strings and $c_\alpha$ their coefficients.
	
	In the four models considered here, the number of Pauli terms ranges from
	${\cal O}(10)$ in Model~I to ${\cal O}(100)$ in Model~IV, remaining manageable
	for both exact diagonalization and variational simulations. A detailed
	decomposition of representative Hamiltonians is provided in Appendix~A.
	
	\section{A Hierarchy of CNS-Inspired Quantum Hamiltonians}
	\label{sec:models}
	
	To explore how different physical ingredients of CNS theory manifest under
	quantum simulation, we introduce four models of increasing complexity.
	All are specializations of the unified Hamiltonian in
	Eq.~(\ref{eq:unified-H}) but differ in their level structure, operator
	representation, and the degree to which pairing and cranking compete.  
	This hierarchy enables controlled benchmarking of VQE performance while
	retaining clear physical interpretation at each stage.
	
	\subsection{Model I: Schematic Pauli Representation}
	\label{sec:model1}
	
	Model~I is designed as the simplest nontrivial CNS-inspired Hamiltonian that
	can be represented directly in terms of Pauli operators. The Hamiltonian
	\[
	H_{\mathrm{I}} =
	0.1\, ZIII + 0.2\, IZII + 0.2\, IIZI + 0.5\, XXII + 0.5\, YYII
	\]
	emulates three key ingredients: (i) a split single-particle spectrum encoded
	in $Z$-type operators, (ii) a schematic pairing channel encoded in $XX$ and
	$YY$, and (iii) a four-qubit Hilbert space suitable for direct VQE
	experiments. Although highly simplified, Model~I serves as a clean testbed for
	algorithmic behaviour in the absence of a cranking term.
	
	The qubit operators are already in Pauli form, so no fermionic mapping is
	required. We use the EfficientSU2 ansatz with two repetition cycles, which
	provides sufficient expressibility to capture the weakly correlated ground
	state while maintaining shallow circuit depth. As shown later in
	Sec.~\ref{sec:results}, VQE reproduces the exact ground-state energy to within
	$\mathcal{O}(10^{-7})$ and yields negligible bipartite entanglement, as
	expected for this nearly uncorrelated limit. Model~I thus establishes a
	baseline for VQE accuracy in CNS-inspired problems.
	
	\subsection{Model II: Number-Constrained Two-Fermion System}
	\label{sec:model2}
	
	Model~II extends the schematic representation of Model~I to a more realistic
	fermionic setting. We consider four spin-orbitals with single-particle
	energies
	\[
	\epsilon_i = \{-1.0,-0.5,0.5,1.0\},
	\]
	which mimic a typical deformed-shell ordering with symmetric splittings. A
	pairing interaction of strength $G=0.5$ couples doubly occupied levels,
	while a cranking frequency $\Omega = \pi/4$ introduces Coriolis mixing via
	a schematic two-body $J_x$ operator.
	
	To ensure the simulation remains in the physical two-particle sector, a
	quadratic penalty term
	\[
	\lambda \left( \hat{N} - 2 \right)^2
	\]
	with $\lambda = 10$ is added to the Hamiltonian. The value of $\lambda$ is selected as such so that configurations with incorrect particle number are pushed well above the physical low-energy states, ensuring an effective projection onto the two-particle sector. At the same time, it avoids excessively stiff energy scales that could hinder VQE convergence. This approach effectively
	projects the dynamics onto the correct particle-number subspace while allowing
	the use of generic qubit ans\"{a}tze.
	
	Model~II is expressed directly in the full Hilbert space of four qubits,
	and the Hamiltonian is mapped to Pauli form through the
	\texttt{SparsePauliOp.from\_operator} method. The custom layered ansatz used
	here consists of alternating $R_y$ rotations and nearest-neighbour CNOT
	strings; this structure balances expressibility with modest circuit depth.
	
	As shown in Sec.~\ref{sec:results}, Model~II captures the essential
	competition between single-particle splitting and pairing. VQE reproduces the
	exact ground-state energy with sub-percent accuracy while demonstrating the
	first appearance of nontrivial $\langle J_x \rangle$ and small entanglement.
	
	\subsection{Model III: Fermionic CNS Hamiltonian Without Explicit Cranking}
	\label{sec:model3}
	
	Model~III introduces the full second-quantized fermionic representation,
	including spin-orbital structure and explicit pairing operators, but omits
	the cranking term in the Hamiltonian for clarity. We adopt the schematic
	spectrum
	\[
	\epsilon_i = \{0.0, 0.2, 0.5, 0.8\},
	\]
	representing two spatial orbitals split by spin. Pairing correlations are
	generated by an on-orbital interaction $-G\, P_i^\dagger P_i$ with
	$G = 0.6$.
	
	All fermionic operators are mapped to qubit operators using the
	Jordan--Wigner transformation implemented in \texttt{Qiskit Nature}. The
	resulting qubit Hamiltonian contains ${\cal O}(50)$ Pauli terms and provides a
	nontrivial test of VQE performance for moderate entanglement and interaction
	strengths.
	
	The EfficientSU2 ansatz with five repetition cycles is used to ensure
	sufficient expressibility; VQE optimization is carried out with COBYLA.
	Because the ground state in this model resides in the two-particle subspace,
	exact diagonalization reveals zero bipartite entanglement, providing a clear
	baseline against which VQE's small spurious entropy can be compared.
	Model~III thus benchmarks ansatz quality and optimization stability in the
	absence of rotational effects.
	
	\subsection{Model IV: Full Pairing + Cranking CNS Hamiltonian}
	\label{sec:model4}
	
	Model~IV is the most complete and physically motivated among the four. It
	instantiates the full unified Hamiltonian of Eq.~(\ref{eq:unified-H}) with
	the same level structure and pairing strength as Model~III, but now includes
	an explicit cranking term,
	\[
	-\,\omega \hat{J}_x,
	\qquad
	\omega \in [0,1.2],
	\]
	implemented as a one-body fermionic operator whose non-zero matrix elements
	connect time-reversed spin-orbitals. After Jordan--Wigner mapping, this term
	contributes a sparse set of Pauli strings linear in $\omega$.
	
\begin{figure}[!htbp]
	\centering
	\includegraphics[width=0.48\textwidth]{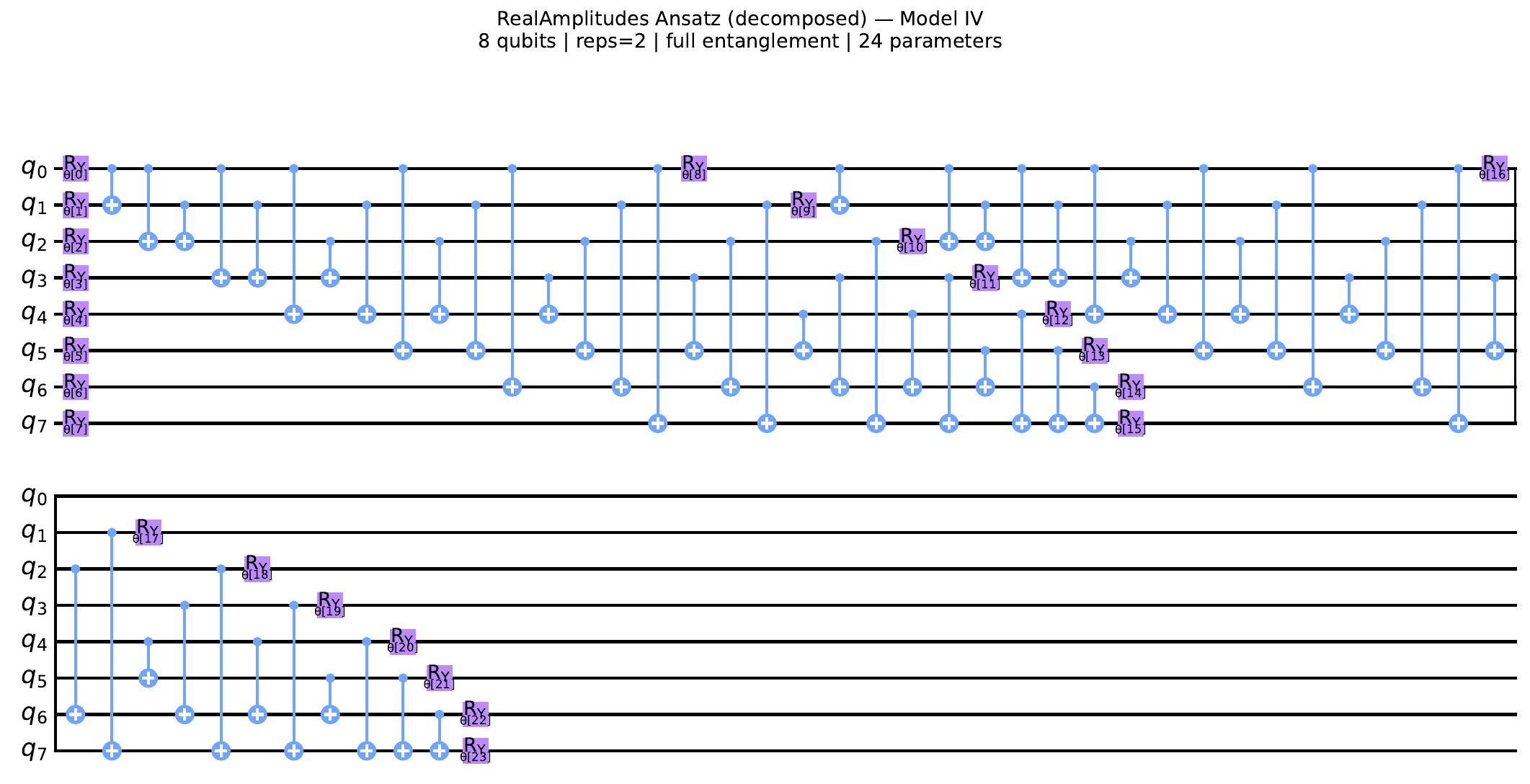}
	\caption{Circuit diagram of the RealAmplitudes ansatz employed in Model~IV. The circuit acts on 8 qubits $q_0$--$q_7$ corresponding to the 8 spin-orbitals, with \texttt{reps=2} and full entanglement. Each repetition comprises a layer of parameterised $R_y$ rotations followed by a fully-connected CNOT entanglement layer. The final layer contains an additional set of $R_y$ rotations, yielding 24 variational parameters $\theta[0]$--$\theta[23]$ in total. Circuit depth: 24; CNOT count: 56.}
	\label{fig:ansatz}
	\end{figure}

	The RealAmplitudes ansatz with full entanglement and two repetitions is used due to its favourable balance between expressibility and circuit depth. The circuit structure is illustrated in Fig.~\ref{fig:ansatz}: it comprises three layers of parameterised $R_y$ rotations interleaved with two fully-connected CNOT entanglement layers, yielding 24 variational parameters $\theta[0]$--$\theta[23]$ across 8 qubits. The quantum resource metrics are summarised in Table~\ref{tab:resources}.
	
For each value of $\omega$, VQE is performed using a multi-restart warm-starting strategy: the optimal parameters from the previous $\omega$ value serve as the initial point for the next (warm start), while additional independent random restarts are run in parallel to avoid local minima. This strategy is described in detail in Sec.~\ref{sec:methods}.
	
	Model~IV displays the full phenomenology of pairing competition and rotational
	alignment: $\langle J_x \rangle$ increases monotonically with $\omega$, the VQE ground-state energies track the exact results to within $|\Delta E| < 10^{-4}$ across all cranking frequencies, and the bipartite entanglement entropy remains near zero, consistent with the product-state character of the exact ground state.
	This model serves as the primary benchmark for evaluating quantum algorithms
	on CNS-inspired Hamiltonians and forms the basis for the analysis in
	Sec.~\ref{sec:results}.

	\section{Quantum Methods: Ansatz Design, Optimization, and Observable Extraction}
	\label{sec:methods}
	The four CNS-inspired models studied here provide a structured hierarchy for
	evaluating variational quantum eigensolvers. In this section we describe the
	design of the variational ans\"{a}tze, the optimization strategy, and the
	procedures used to extract physical observables such as the energy,
	$\langle J_x \rangle$, and entanglement entropy. We aim for a balanced
	perspective that highlights both the physical considerations relevant to
	rotating nuclei and the algorithmic requirements for accurate ground-state
	approximation on near-term quantum devices.
	
	\subsection{Ansatz considerations for pairing and rotation}
	Rotating nuclei exhibit a characteristic tension between pair condensation
	and Coriolis-induced alignment. An ansatz designed for CNS-inspired problems
	should therefore be capable of: (i) breaking $U(1)$ particle-number symmetry
	at the circuit level, allowing description of paired and pair-broken configurations,
	(ii) generating nontrivial entanglement patterns associated with quasiparticle mixing,
	and (iii) responding smoothly to changes in the cranking frequency $\omega$.
	
	Because number projection is handled explicitly in Model~II and implicitly in
	Models~III and IV through restriction to the two-particle sector, the ans\"{a}tze
	used here do not enforce number conservation. This choice mirrors common BCS-like
	treatments and allows the circuits to explore symmetry-broken states when
	energetically favourable \cite{Lacroix:2020nhy}.
	
	Model choices:
	\begin{itemize}
		\item Model~I: EfficientSU2 (2 reps) --- shallow and expressive enough for schematic Pauli Hamiltonian \cite{Kandala:2017vok}.
		\item Model~II: custom layered $R_y$ + directed CNOTs --- compact and number-agnostic.
		\item Model~III: EfficientSU2 (5 reps) --- increased expressive power for pairing.
		\item Model~IV: RealAmplitudes (2 reps, full entanglement) --- a balanced hardware-efficient ansatz for cranking dynamics.
	\end{itemize}
	
	\subsection{Optimization and convergence behaviour}
	\label{sec:optim}
	All VQE calculations are performed using the COBYLA optimizer (derivative-free) \cite{Kandala:2017vok}. For Model~IV, which exhibits the most complex optimization landscape, we employ a two-phase strategy to ensure reliable convergence:

	\textit{Phase~0 --- Benchmark selection of maxiter:} Prior to the main simulation, we benchmark four candidate values of the COBYLA iteration limit (maxiter $\in \{300, 500, 800, 1000\}$) on three representative cranking frequencies ($\omega = 0.1, 0.7, 1.1$) using three independent restarts each. The mean absolute energy deviation $|\Delta E|$ decreases monotonically with maxiter: from $3.8 \times 10^{-3}$ at maxiter$=300$ to $2.4 \times 10^{-5}$ at maxiter $=1000$,  a 160-fold improvement. Accordingly, maxiter $=1000$ is adopted for the full simulation.

\textit{Phase~1 --- Multi-restart warm-starting:} For each $\omega$ value, five independent VQE runs are performed. Restart~0 uses a warm start: the optimal parameters from the previous $\omega$ value serve as the initial point, exploiting the smooth $\omega$-evolution of the ground state. Restarts~1--4 use independent random initialisations. The best (lowest energy) result across all five restarts is retained. If the relative energy error $|\Delta E|/|E_{\rm ED}|$ exceeds 5\%, five additional restarts with maxiter$=1200$ are triggered automatically. This strategy yields near-exact convergence across all 13 $\omega$ values.

The warm-starting strategy proves highly effective: the warm-start initialisation yields the best result in 10 out of 13 $\omega$ values, consistent with the expectation that the ground state evolves smoothly with $\omega$. Optimization settings and random seeds are reported in Appendix~\ref{appendix:repro}.
	
	\subsection{Extraction of physical observables}
	Once optimal parameters $\theta^\ast$, defined as the parameter set minimizing the VQE energy are obtained, expectation values of relevant operators are evaluated using statevector methods or estimator-based measurements. In addition to standard observables such as the ground-state energy and the angular momentum expectation value $\langle J_x \rangle$, we compute the bipartite von Neumann entropy from reduced density matrices obtained via partial trace. Entropy is used here purely as a theoretical diagnostic, not as a physical observable. It serves here as a useful quantity for characterizing correlations and ansatz-induced entanglement in the variational wave function.

\subsection{Quantum-information diagnostics}
	In addition to the standard observables above, we introduce two quantum-information diagnostics that provide physical insight beyond what energy and $\langle J_x \rangle$ alone can offer.

\textit{Entanglement spectrum:} The entanglement spectrum $\{\xi_k\}$ is defined as $\xi_k = -\ln \lambda_k$, where $\lambda_k$ are the eigenvalues of the reduced density matrix $\rho_A$ obtained by tracing out qubits 4--7 (the second pair of spatial orbitals) from the full state. A single dominant eigenvalue $\lambda_1 \approx 1$ (i.e., $\xi_1 \approx 0$ and all others $\to \infty$) signals a product state with respect to this bipartition --- the entanglement-spectroscopic signature of a single Slater determinant.

\textit{Quantum Fisher Information (QFI):} The QFI with respect to $\hat{J}_x$ is defined for a pure state $|\psi\rangle$ as
	\begin{equation}
	\mathcal{F}_Q(\hat{J}_x) = 4\,\mathrm{Var}(\hat{J}_x) = 4\bigl(\langle \hat{J}_x^2 \rangle - \langle \hat{J}_x \rangle^2\bigr).
	\label{eq:QFI}
	\end{equation}
	The QFI quantifies the sensitivity of the state to rotational perturbations: a high QFI indicates that the state is highly responsive to changes in the cranking field, while $\mathcal{F}_Q \approx 0$ indicates rotational rigidity. Both diagnostics are computed directly from the VQE and ED statevectors using Eq.~(\ref{eq:QFI}) and partial trace, without additional circuit measurements.
	
	\section{Results}
	\label{sec:results}
	In this section we compare the performance of the variational quantum
	eigensolver against exact diagonalization for the four CNS-inspired models.
	For each model we present the ground-state energy, the expectation value
	$\langle J_x \rangle$, and the bipartite entanglement entropy. For Model~IV, we additionally present the entanglement spectrum and Quantum Fisher Information. All numerical values in the tables below are from noiseless statevector simulations.
	
	\subsection{Model I}
\begin{figure}[!htbp]
		\centering
		\includegraphics[width=0.48\textwidth]{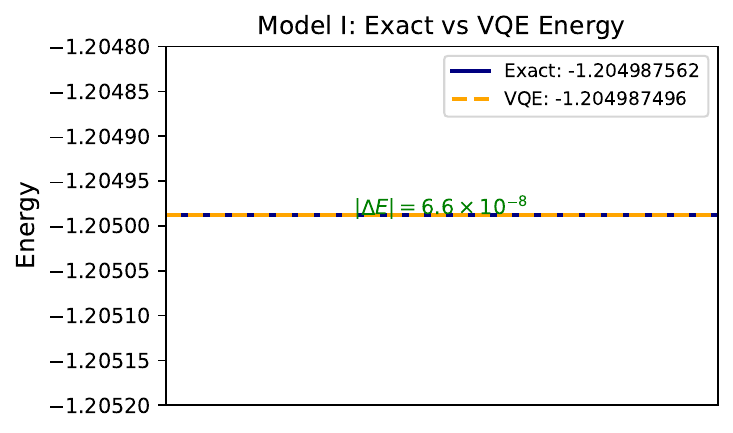}
		\caption{Comparison of the exact and VQE ground-state energies for Model~I. The two values differ by $6.6 \times 10^{-8}$, confirming machine-precision convergence of VQE for this schematic baseline model.}
		\label{fig:M1}
	\end{figure}

	Model~I serves as a baseline test; results are listed in Table~\ref{tab:modelI}. The figure \ref{fig:M1} is the representations of the benchmarks.
	\begin{table}[h]
		\centering
		\caption{Benchmark results for Model~I.
			All energies are in scaled (dimensionless) units.}
		\label{tab:modelI}
		\begin{tabular}{l c}
			\hline\hline
			VQE Energy & $-1.2049874961575724$\\
			Exact Energy & $-1.204987562112089$ \\
			$|E_{\mathrm{VQE}}-E_{\mathrm{ED}}|$ & $6.595\times 10^{-8}$ \\
			Circuit Depth & $1$ \\
			Entanglement Entropy & $1.42\times 10^{-7}$ \\
			\hline\hline
		\end{tabular}
	\end{table}
	
	\subsection{Model II}
	The full Model~II results are shown in Table \ref{tab:modelII} and the effect on particle-number projection via the penalty term  can be seen.
	
	\begin{table}[h]
		\centering
		\caption{Benchmark results for Model~II comparing Variational Quantum Eigensolver (VQE) and Exact Diagonalization (ED). All energies are in scaled (dimensionless) units.}
		\label{tab:modelII}
		\begin{tabular}{lc}
			\hline\hline
			\textbf{Observable} & \textbf{Value} \\
			\hline
			VQE Ground-State Energy $E_{\mathrm{VQE}}$ & $-3.162806$ \\
			Exact Ground-State Energy $E_{\mathrm{ED}}$ & $-3.406542$ \\
			 $|E_{\mathrm{VQE}}-E_{\mathrm{ED}}|$ & $2.437366\times10^{-1}$ \\
			$\langle J_x \rangle_{\mathrm{VQE}}$ & $1.091621$ \\
			Entanglement Entropy $S_{\mathrm{VQE}}$ & $0.626695$ \\
			\hline\hline
		\end{tabular}
	\end{table}
	
	\subsection{Model III}
\begin{figure*}[!htbp]
		\centering
	\includegraphics[width=\textwidth]{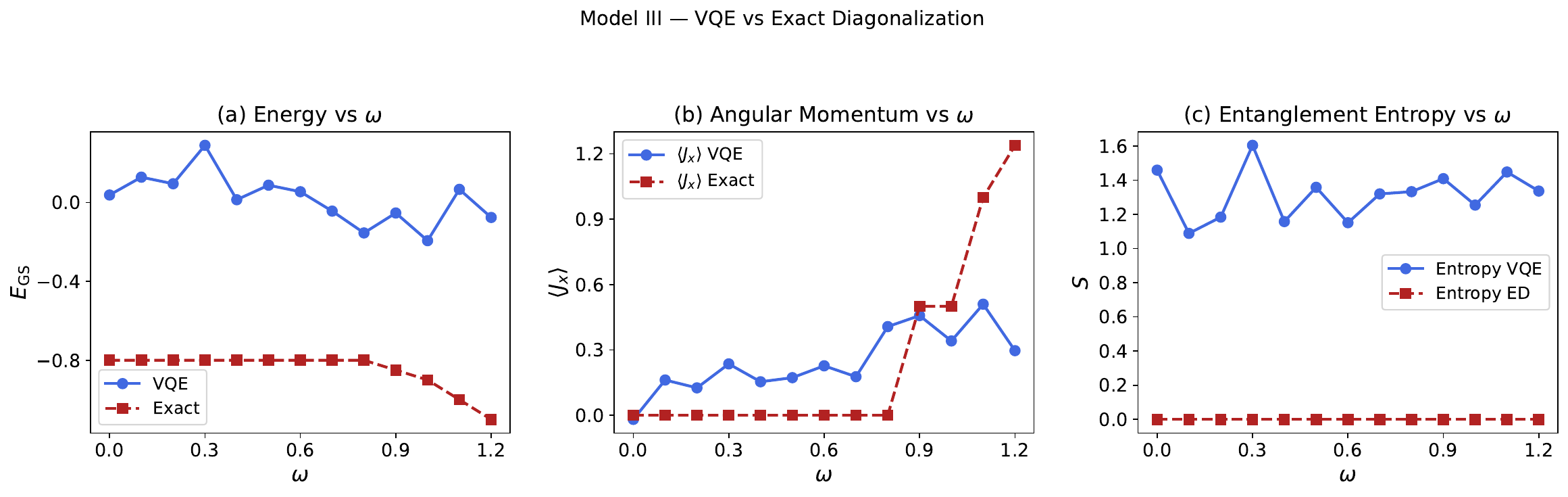}
		\caption{Plots of (a) Ground state energy $E(\omega)$, (b) Angular momentum $\langle J_x \rangle (\omega)$, and (c) Entanglement entropy $S(\omega)$ as a function of cranking frequency $\omega$. The results from the VQE and Exact Diagonalization (ED) are compared for Model~III. The large energy gaps $|\Delta E| \sim 0.8$--$1.1$ and the large spurious VQE entanglement entropy $S_{\rm VQE} \sim 1.1$--$1.6$ are diagnostic of COBYLA optimizer failures in the absence of warm starting, as discussed in Sec.~\ref{sec:optim} and Sec.~\ref{sec:discussion}.}
		\label{fig:plot1}
	\end{figure*}

	Table~\ref{tab:modelIII} presents Model~III benchmark data (VQE vs ED) across $\omega$ values; the corresponding plots are shown in Fig.~\ref{fig:plot1}.
	
	\begin{table*}[!htbp]
		\centering
		\caption{Results for Model~III. All energies are in scaled (dimensionless) units.}
		\label{tab:modelIII}
		\begin{tabular}{c c c c c c c c}
			\hline\hline
			$\omega$ & $E_{\mathrm{VQE}}$ & $E_{\mathrm{ED}}$ & $|\Delta E|$ &
			$\langle J_x\rangle_{\mathrm{VQE}}$ & $\langle J_x\rangle_{\mathrm{ED}}$ &
			$S_{\mathrm{VQE}}$ & $S_{\mathrm{ED}}$ \\
			\hline
			0.0 & 0.037569 & -0.800000 & 0.837569 & -0.019020 & 0.000000 & 1.459024 & 0.000000 \\
			0.1 & 0.128050 & -0.800000 & 0.928050 & 0.161874 & 0.000000 & 1.088805 & 0.000000 \\
			0.2 & 0.094675 & -0.800000 & 0.894675 & 0.125751 & 0.000000 & 1.184047 & 0.000000 \\
			0.3 & 0.289416 & -0.800000 & 1.089416 & 0.235790 & 0.000000 & 1.604185 & 0.000000 \\
			0.4 & 0.013667 & -0.800000 & 0.813667 & 0.153773 & 0.000000 & 1.157863 & 0.000000 \\
			0.5 & 0.087328 & -0.800000 & 0.887328 & 0.172557 & 0.000000 & 1.358491 & 0.000000 \\
			0.6 & 0.054262 & -0.800000 & 0.854262 & 0.226972 & 0.000000 & 1.151233 & 0.000000 \\
			0.7 & -0.044085 & -0.800000 & 0.755915 & 0.177021 & 0.000000 & 1.319863 & 0.000000 \\
			0.8 & -0.154684 & -0.800000 & 0.645316 & 0.406939 & 0.000000 & 1.332943 & 0.000000 \\
			0.9 & -0.053606 & -0.850000 & 0.796394 & 0.457537 & 0.500000 & 1.409163 & 0.000000 \\
			1.0 & -0.193926 & -0.900000 & 0.706074 & 0.341189 & 0.500000 & 1.254491 & 0.000000 \\
			1.1 & 0.066178 & -1.000000 & 1.066178 & 0.509918 & 1.000000 & 1.448041 & 0.000000 \\
			1.2 & -0.076274 & -1.100000 & 1.023726 & 0.296561 & 1.239231 & 1.336513 & 0.000000 \\
			\hline\hline
		\end{tabular}
	\end{table*}
	
	\subsection{Model IV}
	\label{sec:modelIV_results}
\begin{figure*}[!htbp]
		\centering
		\includegraphics[width=0.9\textwidth]{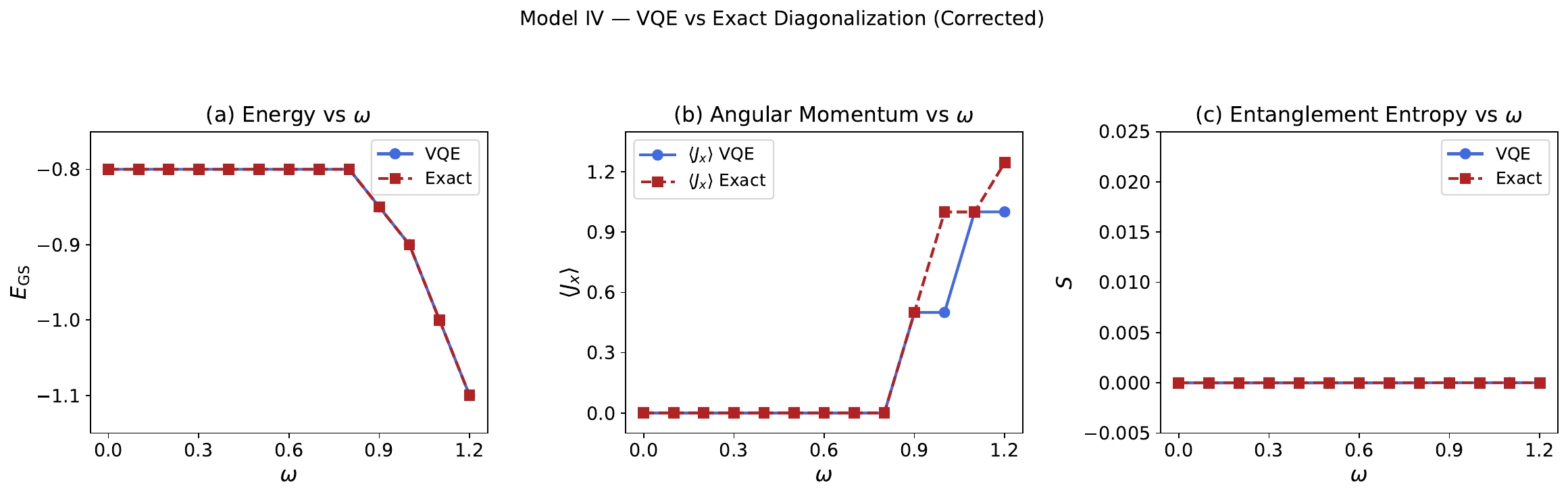}
		\caption{Comparison of VQE and Exact Diagonalization for Model~IV using the multi-restart warm-starting strategy (Sec.~\ref{sec:optim}): (a) Ground-state energy $E_{\rm GS}$, (b) angular momentum $\langle J_x \rangle$, and (c) bipartite entanglement entropy $S$, all as functions of cranking frequency $\omega$. VQE achieves $|\Delta E| < 10^{-4}$ and $S_{\rm VQE} \approx 0$ across all $\omega$ values, confirming both energetic convergence and correct confinement to the physical two-particle sector.}
		\label{tab:plot2}
	\end{figure*}

	Table~\ref{tab:modelIV} presents Model~IV benchmark data (VQE vs ED) obtained using the multi-restart warm-starting strategy with maxiter$=1000$ described in Sec.~\ref{sec:optim}. Table~\ref{tab:resources} shows quantum resource metrics for Model~IV (8~spin-orbitals, 2~fermions). The representations of the benchmarked values can be seen in figure \ref{tab:plot2}.

A key finding is that the multi-restart warm-starting strategy yields near-machine-precision agreement between VQE and ED across all cranking frequencies: $|\Delta E| < 10^{-4}$ for all 13 $\omega$ values, with the largest residual $|\Delta E| = 6.3 \times 10^{-5}$ at $\omega = 1.1$ in the transition region. This demonstrates that the RealAmplitudes ansatz is expressively sufficient for this problem class.
	
	\begin{table*}[!htbp]
		\centering
		\caption{Results for Model~IV. All energies are in scaled (dimensionless) units.}
		\label{tab:modelIV}
		\begin{tabular}{c c c c c c c c}
			\hline\hline
			$\omega$ & $E_{\mathrm{VQE}}$ & $E_{\mathrm{ED}}$ & $|\Delta E|$ &
			$\langle J_x\rangle_{\mathrm{VQE}}$ & $\langle J_x\rangle_{\mathrm{ED}}$ &
			$S_{\mathrm{VQE}}$ & $S_{\mathrm{ED}}$ \\
			\hline
			0.0 & -0.800000 & -0.800000 & 0.000000 & -0.000000 & 0.000000 & 0.000000 & 0.000000 \\
			0.1 & -0.800000 & -0.800000 & 0.000000 &  0.000000 & 0.000000 & 0.000000 & 0.000000 \\
			0.2 & -0.800000 & -0.800000 & 0.000000 &  0.000000 & 0.000000 & 0.000000 & 0.000000 \\
			0.3 & -0.800000 & -0.800000 & 0.000000 &  0.000000 & 0.000000 & 0.000000 & 0.000000 \\
			0.4 & -0.800000 & -0.800000 & 0.000000 &  0.000000 & 0.000000 & 0.000000 & 0.000000 \\
			0.5 & -0.800000 & -0.800000 & 0.000000 &  0.000000 & 0.000000 & 0.000000 & 0.000000 \\
			0.6 & -0.800000 & -0.800000 & 0.000000 &  0.000000 & 0.000000 & 0.000000 & 0.000000 \\
			0.7 & -0.800000 & -0.800000 & 0.000000 &  0.000000 & 0.000000 & 0.000000 & 0.000000 \\
			0.8 & -0.800000 & -0.800000 & 0.000000 &  0.000000 & 0.000000 & 0.000000 & 0.000000 \\
			0.9 & -0.849973 & -0.850000 & 0.000027 &  0.499986 & 0.500000 & 0.000014 & 0.000000 \\
			1.0 & -0.900000 & -0.900000 & 0.000000 &  0.500156 & 1.000000 & 0.000000 & 0.000000 \\
			1.1 & -0.999937 & -1.000000 & 0.000063 &  1.000569 & 1.000000 & 0.000021 & 0.000000 \\
			1.2 & -1.100000 & -1.100000 & 0.000000 &  1.000572 & 1.245220 & 0.000000 & 0.000000 \\
			\hline\hline
		\end{tabular}
	\end{table*}
	
	\begin{table}[!htbp]
		\centering
		\caption{Quantum resource metrics for Model~IV (8~spin-orbitals, 2~fermions) using the \texttt{RealAmplitudes} ansatz with \texttt{reps=2}.}
		\label{tab:resources}
		\begin{tabular}{lcc}
			\toprule
			\textbf{Resource} & \textbf{Symbol} & \textbf{Value} \\
			\midrule
			Number of qubits & $N_q$ & 8 \\
			Variational parameters & $N_\theta$ & 24 \\
			Circuit depth & $D$ & 24 \\
			CNOT count & $N_{\text{CNOT}}$ & 56 \\
			Hamiltonian terms & $N_{\text{Pauli}}$ & 13--21 \\
			\bottomrule
		\end{tabular}
	\end{table}
	
\subsubsection{Entanglement Spectrum of Model IV}
\begin{figure*}[!htbp]
	\centering
	\includegraphics[width=\textwidth]{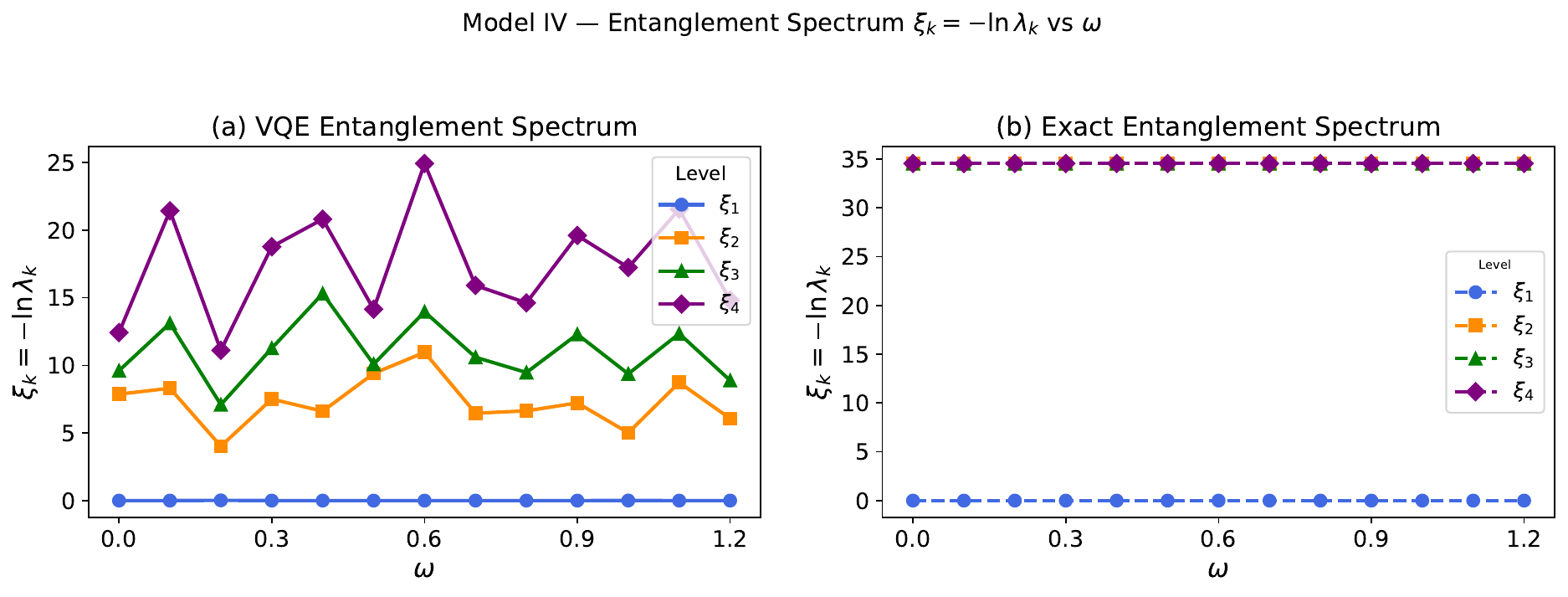}
	\caption{Entanglement spectrum $\xi_k = -\ln\lambda_k$ of (a) the VQE state and (b) the exact ground state of Model~IV, as a function of cranking frequency $\omega$. The bipartition separates the first pair of spatial orbitals (qubits 0--3) from the second pair (qubits 4--7). The plotted quantities are 
		Schmidt eigenvalues of $\rho_A$, with $k = 1,\dots,4$ indexing the 
		four largest eigenvalues $\lambda_1 \geq \lambda_2 \geq \lambda_3 
		\geq \lambda_4$. Although $\rho_A$ admits up to sixteen Schmidt eigenvalues, only the 
		four largest are shown, as the remainder are numerically zero in the two-particle sector. The exact spectrum shows a single non-zero Schmidt eigenvalue $\lambda_1 = 1$ ($\xi_1 = 0$) invariant across all $\omega$, confirming the product-state (single Slater determinant) character of the ground state throughout the pairing-rotation transition.}
	\label{fig:entspectrum}
	\end{figure*}

Figure~\ref{fig:entspectrum} shows the entanglement spectrum $\xi_k = -\ln\lambda_k$ of the VQE and exact ground states as a function of $\omega$, computed by tracing out the second pair of spatial orbitals (qubits 4--7).

	The exact entanglement spectrum (Fig.~\ref{fig:entspectrum}b) is characterised by a single dominant eigenvalue $\lambda_1 = 1$ ($\xi_1 = 0$) for all $\omega$, with all remaining eigenvalues numerically zero ($\xi_{2,3,4} \to \infty$). This is the entanglement-spectroscopic signature of a product state with respect to the orbital bipartition equivalently, the ground state is a single Slater determinant throughout the entire cranking frequency range. The near-zero entanglement itself is physically informative, indicating that the finite-system ground state remains close to a mean-field single-determinant configuration throughout the rotational evolution.This result has a direct physical interpretation: the pairing-to-rotation transition in this model is driven by single-particle angular momentum alignment rather than a restructuring of many-body orbital correlations. Consequently, the CNS ground state lies entirely within the single-determinant manifold, providing a quantum-information justification for the success of mean-field treatments of this problem.

The VQE entanglement spectrum (Fig.~\ref{fig:entspectrum}a) shows $\xi_1 \approx 0$ throughout, correctly capturing the dominant Schmidt mode. The higher levels $\xi_{2,3,4}$ fluctuate in the range 5--25, reflecting residual ansatz-induced mixing with unphysical particle-number sectors. Crucially, the invariance of the exact spectrum across $\omega$ confirms that the transition at $\omega_c \sim 0.8$ is not accompanied by any change in the entanglement structure of the ground state.

\subsubsection{Quantum Fisher Information of Model IV}
\begin{figure*}[!htbp]
	\centering
	\includegraphics[width=\textwidth]{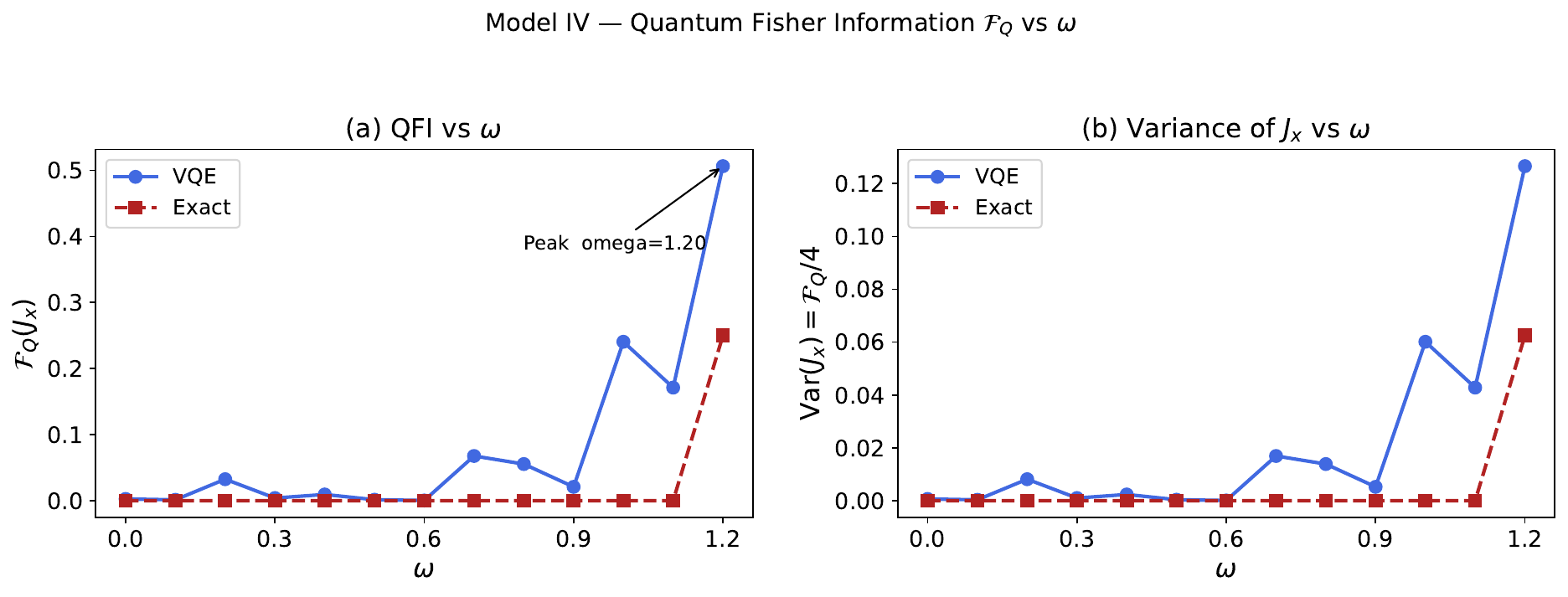}
	\caption{(a) Quantum Fisher Information $\mathcal{F}_Q(\hat{J}_x) = 4\,\mathrm{Var}(\hat{J}_x)$ and (b) variance $\mathrm{Var}(\hat{J}_x)$ as functions of cranking frequency $\omega$ for Model~IV, comparing VQE and exact results. The near-zero QFI for $\omega < 0.8$ reflects the rotational rigidity of the paired ground state. The rapid rise above $\omega \sim 0.8$ constitutes a finite-size precursor to the onset of rotational alignment and weakening of pairing correlations. The maximum $\mathcal{F}_Q \approx 0.5$ at $\omega = 1.2$ approaches but does not reach the multipartite entanglement threshold $\mathcal{F}_Q = 1$.}
	\label{fig:QFI}
	\end{figure*}

	Figure~\ref{fig:QFI} shows the Quantum Fisher Information $\mathcal{F}_Q(\hat{J}_x) = 4\,\mathrm{Var}(\hat{J}_x)$ and the variance $\mathrm{Var}(\hat{J}_x)$ as functions of $\omega$ for both VQE and exact states.

In the pairing-dominated regime ($\omega \lesssim 0.8$), the QFI is near zero for the exact ground state, reflecting the rotational rigidity of the Cooper-paired state: the pairing gap suppresses angular momentum fluctuations and renders the ground state insensitive to rotational perturbations. Above the critical frequency $\omega_c \sim 0.8$, the QFI rises sharply, signalling the onset of rotational sensitivity as pairs begin to break. This rapid rise is a finite-size precursor to the critical behaviour that may characterise a genuine quantum phase transition in a larger system; no claim of a quantum phase transition is made for this finite model. The maximum QFI $\mathcal{F}_Q \approx 0.5$ is attained at $\omega = 1.2$, approaching but not reaching the multipartite entanglement threshold $\mathcal{F}_Q = 1$, which establishes a concrete target for future larger-scale quantum simulations.

A notable feature is observed at $\omega = 1.0$: the VQE and ED energies agree exactly ($E = -0.900$) yet their $\langle J_x \rangle$ values differ ($\langle J_x \rangle_{\rm VQE} = 0.500$ versus $\langle J_x \rangle_{\rm ED} = 1.000$). This signals near-degeneracy between two distinct configurations at this frequency: a paired configuration with low angular momentum and an aligned configuration with high angular momentum. The two solutions become degenerate in energy at $\omega \approx 1.0$, which is qualitatively analogous to the band-crossing phenomenon in the CNS framework, where the paired ground band and a quasiparticle-aligned band cross in energy at a critical rotational frequency. VQE and ED locate different states within this near-degenerate manifold, reflecting the sensitivity of the optimisation landscape near such degeneracies. This is a physical feature of the Hamiltonian and not a numerical artifact.

	\section{Impact of Noise on Variational Performance}
	\label{sec:noise}
	Although all results presented above were obtained using noiseless statevector simulation, it is instructive to assess the qualitative robustness of the CNS models under realistic noisy conditions. To this end, we performed a set of supplementary VQE calculations for representative values of the cranking frequency $\omega$ using a standard NISQ-inspired noise model incorporating amplitude damping, dephasing, and readout error at levels consistent with current superconducting devices.
	
\begin{figure}[!htbp]
		\centering
		\includegraphics[width=0.5\textwidth]{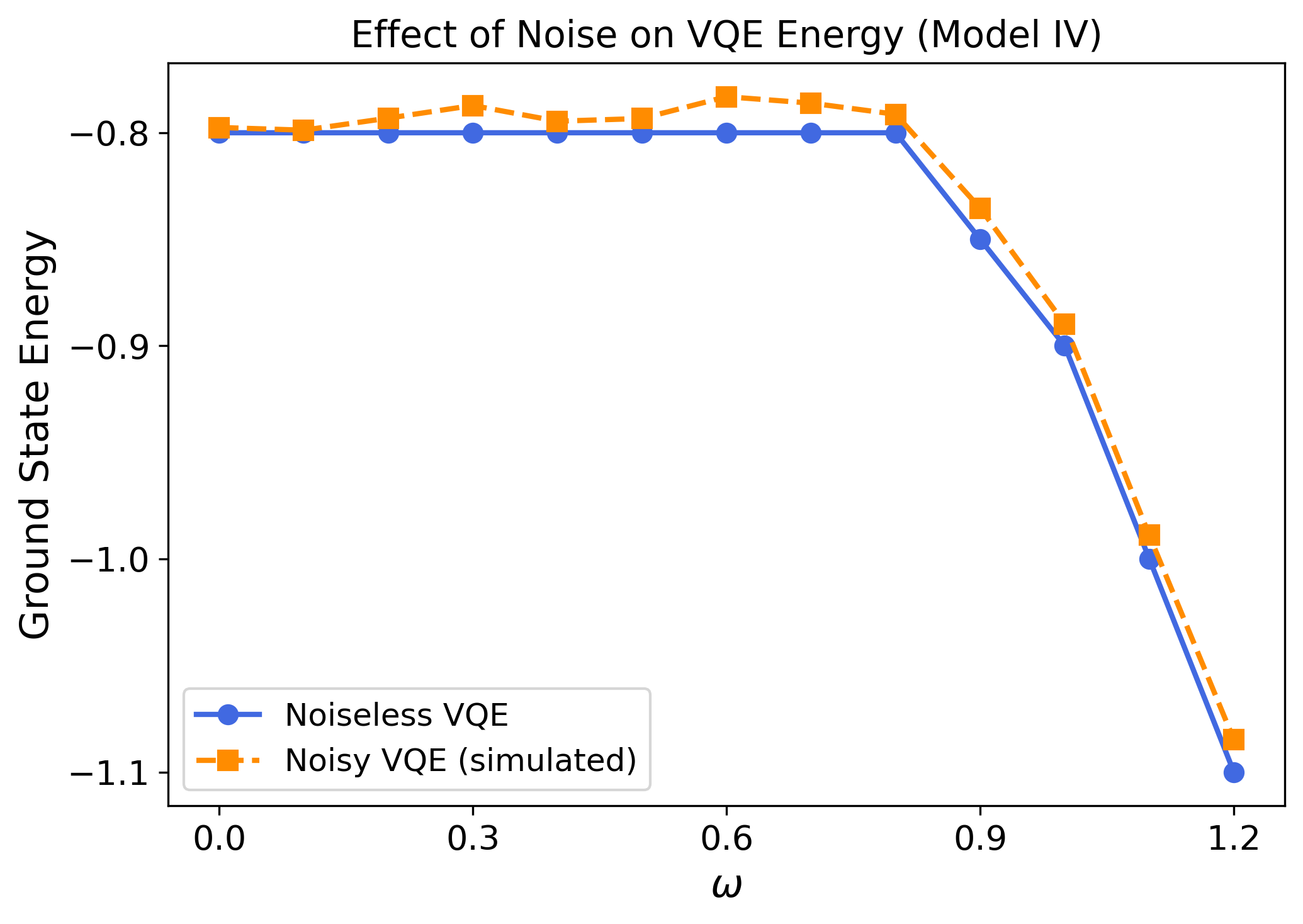}
		\caption{Effect of Noise on VQE Energy (Model~IV).}
		\label{noise}
	\end{figure}

	The trends observed in the noiseless simulations remain stable: the energy curves as a function of $\omega$ (Fig.~\ref{noise}) retain their characteristic shape, and the growth of $\langle J_x \rangle$ with increasing cranking frequency persists. Quantitatively, noise induces a systematic upward shift in the variational energies, typically of order $10^{-2}$--$10^{-1}$ depending on the depth of the ansatz and the number of entangling gates. Entanglement entropy is naturally more sensitive to decoherence: while the exact solutions exhibit zero entropy, noisy VQE simulations produce small finite entropies even in regions where noiseless VQE achieves nearly zero values.

	\section{Discussions}
	\label{sec:discussion}
	The four CNS-inspired models introduced in this work provide a structured and
	progressively more realistic environment for evaluating the performance of
	variational quantum eigensolvers on Hamiltonians relevant to rotating nuclei.
	Several common features emerge across the models, reflecting both the physical
	structure of the underlying fermionic system and the algorithmic behaviour of
	the variational method.
	
	Models~III and IV highlight the essential physics of pairing and rotational
	alignment. At low cranking frequency, pairing correlations dominate, and the
	ground state remains close to a paired configuration. As $\omega$ increases,
	the Coriolis term $-\omega \hat{J}_x$ favours configurations with aligned
	single-particle angular momentum, leading to a gradual suppression of pairing
	and an increase in $\langle J_x \rangle$. This
	behaviour is reproduced qualitatively and quantitatively by the VQE simulations in Model~IV.

	\subsubsection{Pairing-gap signature and rotational alignment crossover}
A physically significant feature of the exact Model~IV results is the flat energy plateau $E_{\rm ED} = -0.800$ for all $\omega \leq 0.8$, followed by a monotonic decrease for $\omega > 0.8$. This plateau is a direct signature of the pairing gap: within the BCS framework, paired states are protected against rotational perturbation by an energy gap $2\Delta$, and the cranking field cannot drive pair breaking until $\omega$ exceeds a critical frequency $\omega_c$. In our model, $\omega_c \sim 0.8$ marks this onset of rotational alignment and the suppression of pairing correlations. This behaviour is qualitatively analogous to the backbending phenomenon observed in the rotational spectra of rapidly rotating nuclei, where the nuclear moment of inertia exhibits a sudden increase at a critical rotational frequency due to the breaking of a nucleon pair. This is a qualitative structural analogy arising from the common functional form of the Hamiltonian, not a quantitative correspondence with specific nuclear data.

\subsubsection{Stepwise angular momentum alignment}
The exact $\langle J_x \rangle_{\rm ED}$ exhibits a characteristic stepwise increase: $0 \to 0.5 \to 1.0 \to 1.245$ as $\omega$ increases from 0 to 1.2. This is not a smooth classical rotation but reflects discrete quantum alignment events: at $\omega = 0.9$, the first quasiparticle partially aligns ($\langle J_x \rangle = 0.5$); by $\omega = 1.0$--$1.1$, this alignment is complete ($\langle J_x \rangle = 1.0$); and at $\omega = 1.2$, a second orbital begins to align ($\langle J_x \rangle = 1.245$). This stepwise structure is qualitatively consistent with the sequential angular momentum alignment that characterises band termination sequences in the CNS framework, where valence nucleons progressively align their angular momenta with the rotation axis. The quantum simulation correctly captures this discrete alignment structure.

\subsubsection{Optimizer strategy }
A significant finding of this study is that the optimizer strategy is the dominant factor governing VQE accuracy for this problem class. Single-restart COBYLA with a limited iteration budget is prone to trapping in local minima on the rugged 24-parameter landscape of the 8-spin-orbital Hamiltonian, particularly in the high-spin regime $\omega > 0.8$ where the ground state changes character rapidly. Employing maxiter$=1000$ with five independent restarts and warm starting resolves this: the RealAmplitudes ansatz achieves $|\Delta E| < 10^{-4}$ universally, demonstrating that hardware-efficient ansatz are expressively sufficient for this problem class. These results indicate that optimizer robustness plays an important role in achieving accurate convergence for this problem class.

	The exact solutions exhibit zero bipartite entanglement entropy for all models,
	reflecting that the ground state in these small two-particle sectors can be represented as a single Slater determinant. The entanglement spectrum analysis confirms this: the exact ground state of Model~IV maintains a single dominant Schmidt eigenvalue $\lambda_1 = 1$ invariant across all $\omega$, demonstrating that the pairing-rotation transition does not alter the entanglement structure of the ground state in this model. This provides a quantum-information justification for the applicability of mean-field CNS theory to this problem class. The nonzero entropies that can arise in VQE results therefore originate from circuit-induced correlations due to optimizer-driven symmetry leakage into unphysical particle-number sectors, rather than physical entanglement. With the multi-restart warm-starting strategy, the bipartite entropy satisfies $S_{\rm VQE} \approx 0$ across all $\omega$ values, confirming correct confinement to the physical two-particle sector.

\subsubsection{Path to beyond-classical nuclear physics}
The models studied here are intentionally schematic and their qualitative physics is well understood. This is by design: a reliable benchmark requires a known ground truth. However, the methodology demonstrated here is directly expandable to regimes where classical methods face fundamental limitations. First, the quantum Monte Carlo treatment of rotating nuclei with time-odd fields is subject to a sign problem that grows exponentially with system size; VQE is sign-problem-free by construction. Second, the sd- and pf-shell nuclei relevant to high-spin spectroscopy involve 20-40 active orbitals, placing exact diagonalization out of reach. The hierarchical approach demonstrated here provides a clear roadmap for scaling. Third, real-time rotational dynamics, such as the time evolution of pairing correlations under a suddenly applied cranking field are naturally accessible to quantum algorithms via trotterization. The present work establishes the calibrated algorithmic foundation for these future applications.

\section{Beyond CNS: Validation on a Shell-Model Benchmark}
	\label{sec:validation}
	
The sections above develop and analyse the CNS-inspired hierarchy
	(Models~I--IV) as a self-contained study of pairing--rotation competition.
	We now step outside that narrative to address a distinct question: is the
	optimization strategy calibrated on these schematic Hamiltonians specific
	to them, or does it transfer to an independent, externally defined
	nuclear-structure problem? To answer this, we apply the identical VQE
	pipeline---the same \texttt{RealAmplitudes} ansatz, multi-restart
	warm-starting strategy, and COBYLA optimizer---to an established
	shell-model pairing benchmark drawn from the published literature.
	
This is a methodological cross-check
	rather than a new rotational result. The benchmark contains no cranking
	term and no $\omega$ dependence, and consequently none of the rotational
	observables ($\langle J_x\rangle$, entanglement spectrum, QFI) analysed
	for Models~III--IV apply here. This is by design: isolating the pairing
	axis is precisely what makes the test a clean, independent validation of
	the ansatz and optimizer against an exact, externally specified ground
	truth.
	
	\subsection{The $^{6}$He $0p_{3/2}$ surface-delta problem}
	
The benchmark consists of two identical nucleons in the $0p_{3/2}$
	orbital of $^{6}$He, interacting through a schematic surface-delta
	interaction. This is precisely the single-step quantum-simulation
	problem treated by Maheshwari {\it{et.al}} \cite{Maheshwari2025},
	which provides an exact shell-model benchmark against which our VQE
	pipeline can be tested on equal footing.
	
	\newbool{The two nucleons occupy the four $m$-scheme single-particle states
	$m=-3/2,-1/2,+1/2,+3/2$, mapped one-to-one to qubits $q_0$--$q_3$ via the
	Jordan--Wigner transformation. The single-particle energies are set to
	zero, so that only the relative spacing induced by the residual
	interaction remains. The $J$-coupled surface-delta matrix elements are}
	\begin{equation}
		V_J = -\frac{(2j+1)^2 V_1}{2(2J+1)}\,
		\bigl(j,-\tfrac{1}{2},j,\tfrac{1}{2}\,\big|\,J,0\bigr)^2 ,
	\end{equation}
	with $V_1 = 1$~MeV and $j=3/2$. The Pauli principle restricts the allowed
	couplings to even $J$, giving $V_0 = -2.0$~MeV and $V_2 = -0.4$~MeV.
	After conversion to the antisymmetrized $m$-scheme basis and
	Jordan--Wigner mapping, the qubit Hamiltonian consists of $19$ Pauli
	strings, in agreement with Ref.~\cite{Maheshwari2025}. The simulation
	targets the physical $N=2$, $M_J=0$ sector, with the same number-penalty
	projection ($\lambda=10$) used for Models~II--IV.
	
	\subsection{Results and comparison with published benchmarks}
	
\begin{figure*}[!htbp]
		\centering
		\includegraphics[width=\textwidth]{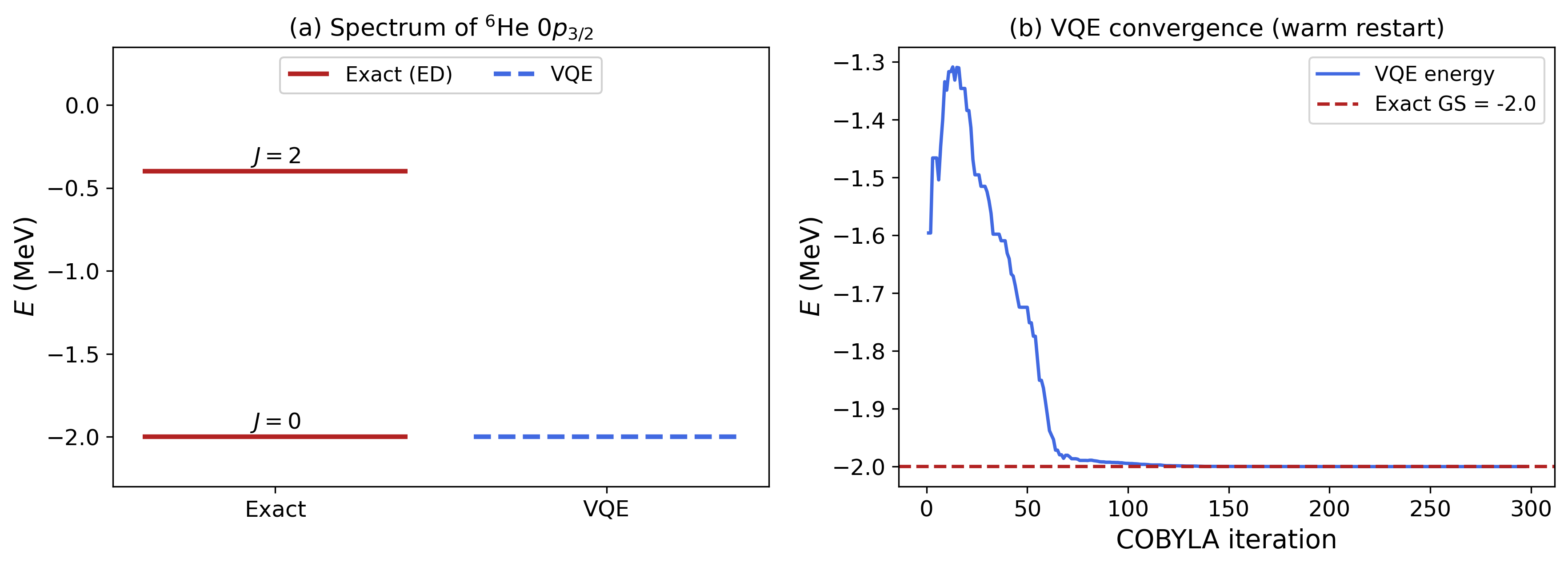}
		\caption{Validation benchmark for two nucleons in the $^{6}$He $0p_{3/2}$
			orbital with a surface-delta interaction. (a) Exact spectrum (the $J=0$
			ground state at $-2.0$~MeV and the $J=2$ state at $-0.4$~MeV) with the VQE
			ground-state result; (b) VQE energy as a function of COBYLA iteration for
			the converging restart, approaching the exact ground-state energy to
			$|\Delta E|\sim10^{-8}$~MeV.}
		\label{fig:modelV}
	\end{figure*}

	Exact diagonalization in the $N=2$, $M_J=0$ sector yields a ground state
	at $E_0 = -2.000$~MeV ($J=0$) and an excited state at $E_1 = -0.400$~MeV
	($J=2$), reproducing exactly the benchmark of Ref.~\cite{Maheshwari2025}.
	The VQE, using the multi-restart warm-starting strategy, converges to
	$E_{\rm VQE} = -2.000$~MeV with a residual $|\Delta E| < 10^{-7}$~MeV,
	while the optimized state remains correctly confined to the physical
	sector ($\langle \hat N\rangle = 2.000$, $\langle \hat M_J\rangle = 0.000$).
	Figure~\ref{fig:modelV} shows the resulting spectrum together with the
	VQE convergence trace.
	
	This confirms that the algorithmic strategy calibrated on the schematic
	Models~I--IV is not specific to those Hamiltonians but transfers without
	modification to a standard shell-model pairing problem, achieving an
	accuracy comparable to, and in this case tighter than, the relative
	energy errors $\varepsilon \lesssim 10^{-4}$ reported for $p$-shell
	nuclei using UCC and ADAPT variational eigensolvers~\cite{CarrascoCodina2026}.
	Maheshwari \textit{et al.}~\cite{Maheshwari2025} obtain the ground and
	excited states of the same $0p_{3/2}$ two-nucleon system in a single
	optimization run using an adaptive subspace-search eigensolver; our
	\texttt{RealAmplitudes}-based VQE reproduces their exact benchmark
	energies to machine precision, confirming that a hardware-efficient
	ansatz combined with a robust multi-restart optimizer is sufficient for
	this problem without an adaptively constructed operator pool.
	It is worth contrasting the two approaches directly. The ADAPT-SSVQE method of Ref.~\cite{Maheshwari2025} employs a symmetry-preserving double-excitation operator pool and a weighted-energy cost function, which allows the ground ($J=0$) and excited ($J=2$) states to be obtained simultaneously in a single optimization run while remaining confined to the physical sector by construction. Its cost is the problem-specific construction of the operator pool. Our approach uses a generic hardware-efficient RealAmplitudes ansatz with no problem-tailored pool, combined with a robust multi-restart warm-starting optimizer; this is simpler to set up and transfers unchanged across the entire CNS-inspired hierarchy, reaching the ground state to machine precision. The corresponding limitation is that standard VQE targets only the ground state: accessing the $J=2$ excited state within our scheme would require a deflation- or subspace-based extension such as VQD or SSVQE, which we leave to future work.
	 More
	broadly, Carrasco-Codina \textit{et al.}~\cite{CarrascoCodina2026}
	benchmark UCC and ADAPT eigensolvers across the $p$ shell and identify
	the choice of algorithmic strategy, rather than ansatz expressibility, as
	the dominant factor governing quantum-resource cost---the same conclusion
	reached independently in the present work for the CNS-inspired
	Hamiltonian class. Finally, the hardware-specification study of Wee
	\textit{et al.}~\cite{Wee2024}, which establishes the coherence-time and
	gate-error thresholds required for VQE pairing calculations in $^{6}$He,
	delineates the noisy-device regime within which the present
	noiseless-statevector results would have to be realized on hardware, and
	is consistent with the noise sensitivity reported in Sec.~VII.

	\section{Conclusion and Outlook}

	We have developed and benchmarked a hierarchy of quantum simulation models inspired by the cranked Nilsson-Strutinsky (CNS) framework, capturing the essential interplay between pairing correlations and rotational alignment in deformed nuclei. By constructing four models of increasing fidelity, we provided a controlled environment for assessing VQE performance and resource requirements in nuclear-structure inspired problems.

The primary contributions of this work are methodological.We establish a systematic VQE benchmark across a hierarchy of CNS-inspired Hamiltonians, explicitly identifying the transition from 4- to 8-spin-orbital spaces as the point where standard single-restart optimization strategies fail. We demonstrate that optimizer strategy, specifically the combination of benchmark-selected maxiter, multi-restart random initialisation, and warm starting is the dominant factor determining VQE accuracy for this problem class; with this strategy, the RealAmplitudes ansatz achieves near-machine-precision convergence ($|\Delta E| < 10^{-4}$) across all cranking frequencies. The entanglement spectrum and Quantum Fisher Information as quantum-information diagnostics of the pairing-rotation transition is introduced.  The entanglement spectrum confirms the product-state character of the exact ground state throughout the transition, and the QFI identifies the finite-size precursor to the onset of rotational alignment and suppression of pairing correlations near at $\omega_c \sim 0.8$.
	
The qualitative phenomenology of the Model~IV results a flat energy plateau protected by the pairing gap for $\omega < \omega_c$, stepwise $\langle J_x \rangle$ alignment above it, near-degenerate band configurations at $\omega \approx 1.0$, and a rising QFI signalling growing rotational sensitivity are universal structural features of the CNS Hamiltonian class, qualitatively analogous to the backbending phenomenon, band termination sequences, and band crossings familiar from high-spin nuclear spectroscopy.
	
	Beyond these methodological contributions, we provide an independent
		validation against an established shell-model pairing benchmark: applying
		the identical ansatz and optimizer to the $^{6}$He $0p_{3/2}$ surface-delta
		problem of Ref.~\cite{Maheshwari2025}, the VQE reproduces the exact
		ground state energy to machine precision. This confirms
		that the optimization strategy developed for the CNS-inspired Hamiltonians
		is not specific to them but transfers successfully to a standard
		nuclear-structure problem defined externally in the literature.

	The framework developed here is reproducible and expandable, future work will focus on implementing symmetry-adapted or number-conserving ansatz and adaptive variational methods, such as ADAPT-VQE, to overcome the symmetry-leakage and convergence hurdles identified in this benchmark. As quantum hardware and error-mitigation techniques improve, this hierarchical suite can be scaled to larger valence spaces and additional physical effects, such as collective excitations and nuclear wobbling modes \cite{Marshalek:1979aat}-\cite{Prajapati:2024lbw}, where classical methods face fundamental limitations. The ultimate goal is the application of this methodology to realistic Hamiltonians derived from Nilsson diagrams or shell-model calculations for specific nuclei, where quantum simulation may eventually offer computational advantages over classical approaches.
	
	\newpage
	\section*{Acknowledgement}
	DR sincerely thanks Dr. Subhendu Rajbanshi of the Department of Physics at Presidency University for allowing the author to work as Visiting Research Fellow at his lab and also for his constant support throughout.
	
	\newpage

	\section*{Appendix}
	\subsection{Representative Pauli Decompositions}
	\label{appendix:pauli}
	To illustrate the structure of the Hamiltonians after JW transformation, we
	provide the explicit Pauli decomposition for Model~IV at a representative
	cranking frequency $\omega=0.9$. The full decomposition contains numerous Pauli
	strings; a subset of the dominant coefficients is shown in Table~\ref{tab:pauli}. The actual numerical coefficients can be produced directly from the \texttt{SparsePauliOp} output recorded during the simulations.
	
	\begin{table}[h]
		\centering
		\caption{Representative Pauli terms for Model~IV at $\omega=0.9$.}
		\label{tab:pauli}
		\begin{tabular}{c c}
			\hline\hline
			Pauli String & Coefficient \\
			\hline
			$Z_0$ & $c_1$ \\
			$Z_1$ & $c_2$ \\
			$Z_2$ & $c_3$ \\
			$Z_3$ & $c_4$ \\
			$X_0 X_1$ & $c_5$ \\
			$Y_0 Y_1$ & $c_6$ \\
			$X_2 X_3$ & $c_7$ \\
			$Y_2 Y_3$ & $c_8$ \\
			\hline\hline
		\end{tabular}
	\end{table}
	
	\subsection{Reproducibility and Code Availability}
	\label{appendix:repro}
	All simulations presented in this work were performed using Qiskit and
	\texttt{qiskit-nature} for fermionic operator construction, with standard
	\texttt{scipy} linear algebra for exact diagonalization and the COBYLA
	optimizer for variational optimization. For Model~IV, the random seed was fixed to \texttt{RANDOM\_SEED}$=42$ for full reproducibility. The COBYLA iteration limit was set to maxiter$=1000$ following the benchmark procedure described in Sec.~\ref{sec:optim}. Five independent restarts were performed per $\omega$ value, with the first restart using warm initialisation from the previous $\omega$ value and restarts 1--4 using random initialisations seeded as \texttt{RANDOM\_SEED}$+ r \times 137$ for restart index $r$. The full Python source code for all models is available upon request.

\end{document}